\documentclass[lettersize,journal]{IEEEtran}
\ifCLASSOPTIONcompsoc
  \usepackage[nocompress]{cite}
\else
  \usepackage{cite}
\fi

\usepackage{algorithm}
\usepackage{algorithmic}
\usepackage{pdfpages}
\usepackage{multirow}
\usepackage[export]{adjustbox} 
\usepackage{bm}
\usepackage{amsmath}
\usepackage{array}
\usepackage{booktabs}
\usepackage{footmisc}
\usepackage{tabularx}

\usepackage{amsthm}
\usepackage{graphicx}
\usepackage{enumitem}
\theoremstyle{definition}
\newtheorem{definition}{Definition}

\usepackage{amsmath,amsfonts,bm}

\def\eqref#1{equation~\ref{#1}}

\def\1{\bm{1}}

\def\ve{{\bm{e}}}

\def\vz{{\bm{z}}}

\def\mE{{\bm{E}}}

\def\mR{{\bm{R}}}

\DeclareMathAlphabet{\mathsfit}{\encodingdefault}{\sfdefault}{m}{sl}
\SetMathAlphabet{\mathsfit}{bold}{\encodingdefault}{\sfdefault}{bx}{n}

\def\gF{{\mathcal{F}}}
\def\gG{{\mathcal{G}}}

\def\gI{{\mathcal{I}}}

\def\gL{{\mathcal{L}}}

\def\gN{{\mathcal{N}}}

\def\gT{{\mathcal{T}}}
\def\gU{{\mathcal{U}}}
\def\gV{{\mathcal{V}}}

\newcommand{\R}{\mathbb{R}}

\usepackage{subfigure}
\usepackage{hyperref}
\usepackage{url}
\usepackage{color}
\usepackage{xcolor}
\usepackage{colortbl}

\begin{document}
\title{Graph Cross-Correlated Network for Recommendation}


\author{Hao Chen\IEEEauthorrefmark{1},
        Yuanchen Bei\IEEEauthorrefmark{1},
        Wenbing Huang,
        Shengyuan Chen,
        Feiran Huang\IEEEauthorrefmark{2},
        and Xiao Huang
\IEEEcompsocitemizethanks{
    \IEEEcompsocthanksitem Hao Chen is with the Faculty of Data Science, City University of Macau, Macao SAR, China. E-mail: sundaychenhao@gmail.com
    \IEEEcompsocthanksitem Yuanchen Bei is with the College of Computer Science and Technology, Zhejiang University, Hangzhou, China. E-mail: yuanchenbei@zju.edu.cn.
    \IEEEcompsocthanksitem Wenbing Huang is with the Gaoling School of Artificial Intelligence, Renmin University of China, Beijing, China. E-mail: hwenbing@126.com.
    \IEEEcompsocthanksitem Shengyuan Chen and Xiao Huang are with the Department of Computing, The Hong Kong Polytechnic University, Hong Kong SAR, China. E-mail: shengyuan.chen@connect.polyu.hk; xiaohuang@comp.polyu.edu.hk.
    \IEEEcompsocthanksitem Feiran Huang is with the College of Information Science and Technology, Guangzhou, China. E-mail: huangfr@jnu.edu.cn.
    \IEEEcompsocthanksitem $*$ Both authors contributed equally to this research.
    \IEEEcompsocthanksitem $\dagger$ Corresponding author: Feiran Huang (huangfr@jnu.edu.cn).
    }
}

\markboth{IEEE TRANSACTIONS ON KNOWLEDGE AND DATA ENGINEERING}
{Shell \MakeLowercase{\textit{et al.}}: Bare Demo of IEEEtran.cls for Computer Society Journals}

%
%

\maketitle
\begin{abstract}
Collaborative filtering (CF) models have demonstrated remarkable performance in recommender systems, which represent users and items as embedding vectors. Recently, due to the powerful modeling capability of graph neural networks for user-item interaction graphs, graph-based CF models have gained increasing attention. They encode each user/item and its subgraph into a single super vector by combining graph embeddings after each graph convolution. However, each hop of the neighbor in the user-item subgraphs carries a specific semantic meaning. Encoding all subgraph information into single vectors and inferring user-item relations with dot products can weaken the semantic information between user and item subgraphs, thus leaving untapped potential.
Exploiting this untapped potential provides insight into improving performance for existing recommendation models. To this end, we propose the Graph Cross-correlated Network for Recommendation (GCR), which serves as a general recommendation paradigm that explicitly considers correlations between user/item subgraphs. GCR first introduces the Plain Graph Representation (PGR) to extract information directly from each hop of neighbors into corresponding PGR vectors. Then, GCR develops Cross-Correlated Aggregation (CCA) to construct possible cross-correlated terms between PGR vectors of user/item subgraphs. Finally, GCR comprehensively incorporates the cross-correlated terms for recommendations. Experimental results show that GCR outperforms state-of-the-art models on both interaction prediction and click-through rate prediction tasks.

\end{abstract}


%

\section{Introduction}\label{sec:introduction}

Collaborative filtering (CF) models have delivered remarkable recommendation performance by representing each user and item as embedding vectors,  where the similarity between users/items is reflected in their embeddings~\cite{he2017ncf,xu2023causal}. Consequently, an intuitive paradigm is to infer the potential relation between users and items by assessing the correlation of their embedding vectors, such as through the dot product or multilayer perceptrons. A typical way to obtain these embeddings is to directly apply the matrix factorization on the user-item interaction matrix~\cite{koren2009MF,rendle2012bprmf}. Moreover,  advancements involve utilizing neural networks to learn interactions~\cite{he2017ncf,tay2018latent} or incorporating additional regularizers~\cite{rao2015grmf,yang2022hrcf}. As research progresses, the question of how to learn powerful and meaningful user/item embeddings has gained significant interest from both the academic and industrial communities.

Recently, there has been a surge in CF-based recommendation models that harness the powerful modeling capabilities of Graph Neural Networks (GNNs) on graph-structured data.
These models leverage graph convolutional processes to extract meaningful graph information from user-item behavior graphs progressively gathering information from distant neighbors within the user/item subgraphs~\cite{wang2019ngcf,he2020lightgcn,wu2022graph}.
Embedding representations for each user and item are derived after each round of graph convolution, converging into a single super vector across layers that encapsulates their interaction patterns.
Subsequently, they follow and employ the conventional approach of computing the dot product of the super vectors to infer the relationship between a given user-item pair.
Representatively, NGCF~\cite{wang2019ngcf} utilizes a concatenation operation to construct the super vectors, while LightGCN~\cite{he2020lightgcn} employs a weighted sum. 
This research direction has consistently demonstrated state-of-the-art performance~\cite{he2020lightgcn,gao2023survey,sharma2024survey}.

\begin{figure*}[thbp]
    \centering
    \includegraphics[width=0.9\linewidth, trim=0cm 0cm 0cm 0cm,clip]{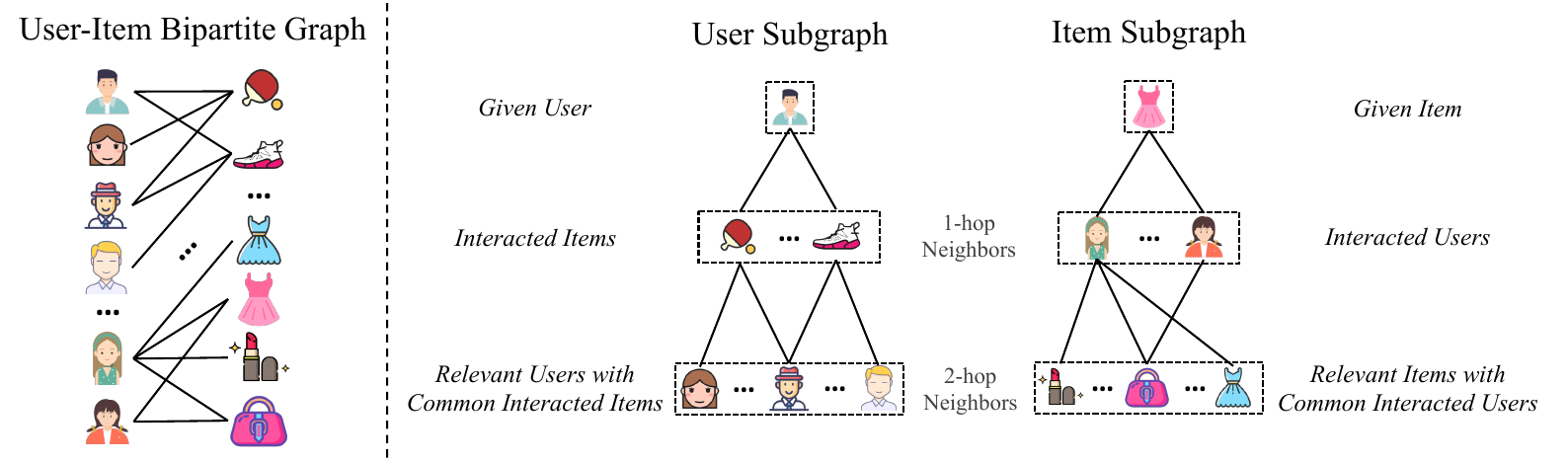}
    \caption{An illustration of the user-item graph and the different semantic meanings of each hop of neighbors in the user/item subgraph.}
    \label{fig:intro}
    \vspace{-0.6em}
\end{figure*}

However, as shown in Figure~\ref{fig:intro}, the recommendation graphs constructed from historical user-item interactions are \textbf{heterogeneous bipartite graphs}~\cite{cao2021bipartite}, where \textit{each hop of neighbor in the user-item subgraphs carries a specific semantic meaning}. For instance, the first-hop neighbors of a given user comprise its historical interacted items, and the second-hop neighbors are relevant users with similar interests who have interacted with common items. Therefore, explicitly considering whether a user has interacted with similar items or whether its relevant users are interested in the target item can provide valuable information. Simply encoding all subgraph information into single vectors and inferring the user-item relation through dot products will undermine and weaken the semantic information between the user and item subgraphs, thereby wasting untapped potential.

By exploiting these potentials, we can not only develop advanced recommendation models but also improve the performance of existing CF models.
Moreover, current GNNs adopt recursive graph convolution to aggregate neighbor embeddings layer by layer~\cite{wang2019ngcf,he2020lightgcn,kipf2016gcn}, resulting in an \textbf{indistinguishable combination} of information from different neighbor hops. 
This blending of information results in impure embeddings for each hop of neighboring nodes.
Based on these observations, the traditional paradigm of using a single vector representation holds untapped potential in improving recommendation performance. Therefore, it is promising to provide non-recursive plain graph representation methods to obtain pure embeddings for each hop of neighbors, enabling the representation of a given user/item subgraph while explicitly extracting and considering all cross-correlations between user/item subgraphs.

Although showing promising prospects, it remains the following challenges:
First, \textbf{lacking suitable paradigms}. Due to the graph convolution and single user-item vector inference architecture in current models, it is not applicable and non-trivial to extract pure embeddings for each hop of neighbors and comprehensively capture meaningful correlations between these embeddings in existing model paradigms.
Moreover, \textbf{ensuring design universality.} Recommendation models in different stages of the recommender system exhibit various characteristics. For instance, the models used in the system's ranking stage and prediction stage have different model architectures and computational efficiency requirements. Consequently, the designed model should adapt to the mainstream recommendation tasks and can plugin into current recommender models.

To this end, in this paper, we investigate how to explore a general graph-based recommendation paradigm to exploit the full potential of embedding vectors in user subgraphs and item subgraphs. Specifically, we aim to investigate the following three research questions: 
(1) How can we obtain pure embedding vectors for each hop of neighbors and represent the given user/item subgraph with more than a single vector?
(2) How to construct and model meaningful cross-correlation terms between the vectors of user subgraphs and item subgraphs?
(3) How to design a flexible paradigm for different recommendation tasks and can be equipped with current recommenders?
By studying these research questions, The main contributions of this paper are organized as follows:
\begin{itemize}
    \item We propose a general graph-based recommendation paradigm called \underline{G}raph \underline{C}rossed-correlated Network for \underline{R}ecommendation (\textbf{GCR}), which represents user/item subgraphs as multiple vectors and explicitly extracts and comprehensively considers the cross-correlation between a pair of user-item subgraphs. GCR is not only an advanced recommendation model but also provides the flexibility to improve the performance of existing recommendation models.
    \item We develop a novel non-recursive plain graph representation architecture for subgraph representation in GCR, called \underline{P}lain \underline{G}raph \underline{R}epresentation (\textbf{PGR}). Different from the previous graph convolution process in the recommendation, PGR first unfolds all neighbors and then aggregates each hop of neighbors, respectively.
    \item We further provide \underline{C}ross-\underline{C}orrelated \underline{A}ggregation (\textbf{CCA}) to explicitly extract and comprehensively consider all cross-correlations between the vectors of the hops of user and item subgraphs. We also theoretically illustrate that GCR has significantly higher flexibility than existing state-of-the-art graph-based recommendation models.
    \item Extensive experimental results on both public datasets and the industrial dataset demonstrate that GCR significantly outperforms the state-of-the-art models. Moreover, replacing the dot product or MLP layer in baseline models with GCR can also significantly improve the recommendation performance of both interaction prediction and click-through rate prediction tasks.
\end{itemize}

\section{RELATED WORK}\label{sec:related}

Current recommendation models can be categorized into three classes based on how the embedding is inferred: Collaborative Filtering (CF), Graph Neural Network (GNN)-based models, and Click-Through Rate (CTR) prediction models. Here, we discuss these types of methods and highlight the necessity of the graph cross-correlated network.

\subsection{Collaborative Filtering}
Collaborative Filtering (CF) is a prevalent technique for recommender systems that parameterizes users and items as embeddings and learns their embeddings based on historical user-item interactions~\cite{he2017ncf,zhang2024multi,hong2024next,chen2021autodebias}. 

One typical type of pioneering work, Matrix Factorization (MF)~\cite{koren2009MF,rendle2012bprmf}, projects the user and item indexes into embedding vectors. It reconstructs historical user-item interactions by computing the dot product between user embeddings and item embeddings.
To improve the quality of these embeddings, researchers have explored incorporating various types of auxiliary information, such as content information~\cite{chen2017attentive,wang2015collaborative}, social relations~\cite{xin2019relational,ning2024cheatagent}, user reviews~\cite{cheng2018aspect,huang2024large}, and knowledge graphs~\cite{wang2019kgat,zhang2023knowgpt,dong2024cost,hong2024knowledge}. Although the dot product can extract interaction information by forcing interacted user-item pairs to achieve a larger dot product value, the linear interaction modeling limits the modeling performance when dealing with complex, sparse, or implicit interactions~\cite{he2017ncf,hsieh2017collaborative}. To enhance the expressive power of the models and capture more intricate interactions, researchers have explored the use of neural networks to learn the interactions~\cite{he2017ncf,zhou2018din}.
This approach offers a promising avenue to improve the approximation capabilities and handle complex patterns in user-item interactions.

Another line of research finds that user-item interactions form a bipartite or heterogeneous graph. In this sense, conventional CF approaches derive user/item embeddings by only considering 1-hop interactions. Therefore, graph embedding methods, initially introduced for social network data mining tasks such as node classification, clustering, and community detection~\cite{tang2015line,huang2017label}, have demonstrated significant potential in real-world recommender systems.
For example, DeepWalk~\cite{perozzi2014deepwalk} and Node2Vec~\cite{grover2016node2vec} adapt word embedding methods~\cite{mikolov2013word2vec} to graph structures and construct the corpus by randomly walking on social networks. Inspired by this, MetaPath2Vec~\cite{dong2017metapath2vec} introduces the metapath-based random walk upon heterogeneous networks to embed users, items, and any other type of nodes together.

\subsection{Graph Neural Networks for Recommendations}
In recent years, graph neural networks (GNNs) have garnered significant attention and have demonstrated state-of-the-art performance across various graph-related tasks~\cite{kipf2016gcn,lagcn,gunets,bei2023reinforcement,chen2023bias,zhao2022popularity}.
Recognizing the strengths of GNNs, researchers have begun incorporating them into Collaborative Filtering (CF) methods to enhance the learning of more powerful embeddings~\cite{he2020lightgcn,chen2024feedback}. 
For instance, 
NGCF~\cite{wang2019ngcf} adapts user-to-item propagation and user-to-user propagation to extract graph embeddings for each user and item and then computes the dot product of concatenated layer-wise graph embeddings. To speed up the training process, LightGCN~\cite{he2020lightgcn} discards the non-linear layers in NGCF and aggregates the layer-wise graph embeddings with weighted summations.
UltraGCN~\cite{mao2021ultragcn} replaces the complex graph convolution process with simple auxiliary objective losses.
Some recent studies further include data augmentation and self-supervised learning for enhancing graph-based recommendation models~\cite{gao2023survey,yu2023self}.
Representatively, SGL~\cite{wu2021sgl} first introduces self-supervised learning on the user-item graph by creating multiple views of a node. It is trained using contrastive learning that maximizes agreement between these different views of the same node while minimizing agreement with views of different nodes. NCL~\cite{lin2022ncl} enhances the neighbor set with semantic neighbors guided by contrastive learning. SimGCL~\cite{yu2022simgcl} is a simple but effective graph contrastive learning model with the removal of unnecessary augmentations. CGCL~\cite{he2023cgcl} further includes the novel strategies of candidate contrastive learning and candidate structural neighbor contrastive learning.
Ultimately, these GNN-based models infer user-item interactions by computing the dot product of the super representation embedding vectors for users and items. By combining the power of GNNs with CF methods, these models leverage the graph structure of user-item interactions~\cite{wu2022graph}.

Although these methods have achieved state-of-the-art recommendation performance, the architecture that simply aggregates the information of users/items and their neighbors might not have enough degrees of freedom to model sophisticated user-item relationships. 
Despite the progress in learning more powerful embeddings, most of the aforementioned methods compute user-item relevance through a simple inner product~\cite{koren2009MF,rendle2012bprmf,he2020lightgcn} or MLP~\cite{he2017ncf,tay2018latent}. Earily, NIA-GCN~\cite{sun2020niagcn} conducts interactions between each central node and its neighbors before aggregation. However, this process does not consider the relevance of the user subgraph and the item subgraph.
There have been few efforts to design more expressive relevance metrics or to investigate how the relevance metrics might influence eventual recommendations.

Therefore, in this paper, our proposed Graph Crossed-correlated Network for Recommendation (GCR) explores whether there exists a learnable relevance metric that can be compatible with any kind of embeddings, including CF embeddings, graph embeddings, and GNN embeddings. This not only helps to enhance the model in existing recommender systems but also enables researchers to gain a deeper understanding of recommendation systems.

\subsection{Click-Through Rate Prediction}
Click-Through Rate (CTR) prediction is a crucial stage for online recommender systems and computational advertisements~\cite{zhou2018din,zhou2019dien,chen2024macro}.
Existing models for CTR
prediction can be largely divided into two main categories: feature-interaction methods and user-interest modeling methods.

Due to the high sparsity of raw input features and the difficulty of directly utilizing such features, \textit{feature interaction-based models} have received extensive research for CTR prediction.
Representatively, FM~\cite{rendle2010factorization} first introduces the latent vectors for 2-order feature interaction to address the feature sparsity. Wide\&Deep~\cite{cheng2016wide} conducts feature interaction by a wide linear regression model and a deep feed-forward network with joint training. 
AFM~\cite{xiao2017attentional} then design the attention mechanism for cross-feature learning.
DeepFM~\cite{guo2017deepfm} further replaces the linear regression in Wide\&Deep with FM to avoid feature engineering. 
Since the user’s historical interactions can reflect the interest, another line of work in recent advances focuses more on \textit{user interest modeling} using deep neural networks and achieves state-of-the-art results. Notably, DIN~\cite{zhou2018din} first designs a deep interest network with an attention mechanism between the user’s behavior sequence and the target item. DIEN~\cite{zhou2019dien} then further enhances DIN with gated recurrent neural units for user’s evolution patterns mining. 
UBR4CTR~\cite{qin2020ubr4ctr} and SIM~\cite{pi2020search} then design a two-stage paradigm, searching relevant items and computing their attention score with the target, to learn of the user’s life-long behaviors. DCIN~\cite{li2023dcin} is a further novel model that integrates explicit and implicit decision-making contexts.
Moreover, GMT~\cite{min2022gmt} and NRCGI~\cite{bei2023non} then include graph sampling and graph clustering for CTR prediction with graph information inclusion in an efficient way.

We have also conducted experimental analyses of equipping the CTR prediction models with our proposed GCR on both the public dataset and the real-world industry dataset, which show that GCR can also be generalized to provide remarkable improvement on the CTR prediction task.

\section{Preliminaries}
{\bf Notations}. We suppose the sets of users and items are denoted as $\gU$ and $\gI$, respectively. 
Then, we use $u\in\gU$ as the user index and $i\in\gI$ as the item index. If no distinction is needed between users and items, we will adopt $j$ as the index. Let $\mR\in\R^{|\gU|\times|\gI|}$ be the user-item interaction matrix, for each $r_{ui}\in \mR$, we have $r_{ui} = y_{ui}$, where $y_{ui}=1$ if the interaction between user $u$ and item $i$ is observed, and otherwise, $y_{ui}=0$. Thus, the bipartite graph is defined as $\gG=(\gU,\gI,\mR)$ implied by the interaction matrix.

We now provide the formal formulation of the recommendation task, followed by introducing several representative methods along with their corresponding mathematical formalization, i.e., CF-based models, GNN-based models, and CTR prediction models.

\begin{definition}[Recommendation Task]
    Given the interaction graph $\gG$, the recommendation task involves predicting the relationships between target users and target items. In form, a recommender model is required to seek to predict these relations with the aim to learn a relevance function $r: \gU\times\gV\rightarrow \R$ based on the interactions in $\gG$, namely, $\hat{y}_{ui} = r(u,i)$ to compute the relevance between $u$ and $i$.
\end{definition}

{\bf CF-based models}. As shown in~\autoref{fig:framework}-(a), CF-based recommender models~\cite{koren2009MF,rao2015grmf,rendle2012bprmf} learn the embeddings of users/items by fitting the constructed interaction matrix $\mR$:
\begin{equation}
    \bm{E}_{\gU}, \bm{E}_{\gI} = f_{CF}(\mR),
\end{equation}
where $\bm{E}_{\gU}$ and $\bm{E}_{\gI}$ denote the learned embedding of the user set and the item set, respectively. 
Then, the inference of the user-item interactions for CF-based recommendation models can be written as:
\begin{equation}
\label{eq:inner-product}
r_{CF}(u,i) = \ve_u^\top \cdot \ve_i,
\end{equation}
where $\ve_u = \bm{E}_{\gU}[u,:] \in\R^d$ and $\ve_i = \bm{E}_{\gI}[i, :]\in\R^d$ are the embeddings of user $u$ and item $i$, respectively, and $d$ denotes the dimension of the embeddings.
By denoting the embeddings as $\mE_{\gU}$ for all users and as $\mE_{\gI}$ for all items, the relevance function in Eq.~(\ref{eq:inner-product}) is generalized as
\begin{equation}
\label{eq:gen}
r_{CF}(u,i) = g(u, i;\gG, \mE_{\gU},\mE_{\gI}),
\end{equation}
where the function $g(\cdot)$ is conditional on the interaction graph and the embeddings.

{\bf GNN-based models}. Graph neural networks~\cite{kipf2016gcn,lagcn} have become one of the most successful tools for graph modeling. When GNNs are tailored for recommendation tasks, a new family of methods is developed~\cite{wang2019ngcf,he2020lightgcn}. As depicted in~\autoref{fig:framework}-(b), GNN-based recommendation models first generate multi-hop graph embeddings by recursively passing messages among neighboring nodes and then aggregate the multi-hop graph embeddings to construct a super vector for the final inference. 

The recursive computation to obtain the multi-hop graph embeddings can be formally written as:
\begin{equation}
\label{eq:GNE}
\ve_j^{(l+1)}=\text{GCL}(\ve_j^{(l)},{\ve_k^{(l)}\ \text{for}\ k\in\gN_j}),
\end{equation}
where $\gN_j$ is defined as the neighbors of node $j$, $\ve_k^{(0)}$ denotes the embedding of node $k$ before the graph convolutional process. $\text{GCL}(\cdot)$ denotes a layer of graph convolution computation process. Both the embeddings and the parameters of GNN are learned from the observed interactions. Some recent state-of-the-art papers, such as LightGCN~\cite{he2020lightgcn}, find that a more lightweight GNN model will be more beneficial.

After obtaining multi-hop graph embeddings, GNN-based models aggregate these embeddings into super vectors with concatenation or weighted summation and then use the dot product of the super vectors to infer user-item interactions. 
Specifically, the representative NGCF~\cite{wang2019ngcf} employs concatenation to aggregate the multi-hop graph embeddings, and the inference of the user-item interactions can be given as:
\begin{equation}
    \begin{aligned}
    r_{NGCF}(u,i) &= (\parallel_{l=0}^L\ve_u^{(l)})^{\top} \cdot (\parallel_{l=0}^L\ve_i^{(l)}) \\
    &= \sum_{l=0}^L\ve_u^{(l)\top} \cdot \ve_i^{(l)},
    \label{eq:NGCF}
    \end{aligned}
\end{equation}
where $L$ denotes the number of graph convolutional layers. Similarly, for the state-of-the-art model LightGCN~\cite{he2020lightgcn}, which aggregates multi-hop graph embeddings with weighted summation, the inference of the user-item interactions can be written as follows:
\begin{equation}
\begin{aligned}
    \ve_x = \sum_{l=0}^Lw_x^{(l)}\ve_x^{(l)},\quad x \in \{u, &i\},\\
    r_{LightGCN}(u, i) = \ve_u ^\top \cdot \ve_i&,
\end{aligned}
\label{eq:LGN}
\end{equation}
where $w_u^{(l)}$ denotes the aggregation weight for user's $l$-hop graph embedding, and $w_i^{(l)}$ denotes the aggregation weight for item's $l$-hop graph embedding.

{\bf CTR prediction models}.
CTR prediction plays a central role in recommendation systems and online advertisements due to its huge commercial value by predicting user preferences towards items/advertisements in real-time~\cite{guo2017deepfm,zhou2018din,bei2023non}.
Our proposed GCR is also flexible to enhance the CTR prediction performance, thus here we provide the formalization of the CTR prediction models.

Given a target user-item pair $(u, i)$, the optional context features $\bm{s}_{ui}$ and the interaction graph $\gG$, the CTR prediction task is to predict the target user $u$'s clicking probability $\hat{y}_{ui}$ on the target item $i$.
In form, the aim of a CTR model is to learn an accurate prediction function $\gF(\cdot)$ with the output of the predicted clicking probability $\hat{y}_{ui}$ as:
\begin{equation}
    \hat{y}_{ui} = \gF(\ve_{u}, \ve_{i}, \bm{s}_{ui}, \gG),
\end{equation}
where the learnable function $\gF(\cdot)$ in CTR prediction models has the objection to minimize the difference from the predicted $\hat{y}_{ui}$ to the ground-truth $y_{ui}$.

Essential notations and definitions we used in this paper can be found in Table~\ref{tab:notations}.

\begin{table}
\caption{Main symbols and definitions in the paper.}
    \begin{tabularx}{\columnwidth}{@{}p{0.2\columnwidth}X@{}}
    \toprule
    Symbol & Description \\
    \midrule
    $\gU$ & the set of users \\
    $\gI$ & the set of items \\
    $\mR$ & the user-item interaction matrix \\
    $\gG=(\gU, \gI, \mR)$ & the user-item bipartite interaction graph\\
    $r_{ui}\in \mR$ & the ground-truth interaction relationship between user $u$ and item $i$ \\
    $d$ & dimension of user/item embeddings \\
    $\mE_{\gU}$ & embedding matrix of the user set $\in \mathbb{R}^{|\gU|\times d}$ \\
    $\mE_{\gI}$ & embedding matrix of the item set $\in \mathbb{R}^{|\gI|\times d}$ \\
    $\ve_u$ & the embedding of the given user $u$ \\
    $\ve_i$ & the embedding of the given item $i$ \\
    $\gN_{j}$ & the neighboring nodes of a user/item node $j$ on the interaction graph \\
    $\hat{y}_{ui}$ & the predicted probability by recommender models that user $u$ will interacted with item $i$ \\
    \bottomrule
    \end{tabularx}
    \label{tab:notations}
\end{table}

\section{Methodology}

\begin{figure*}[t]
	\includegraphics[width=\linewidth, trim=0cm 0cm 0cm 0cm,clip]{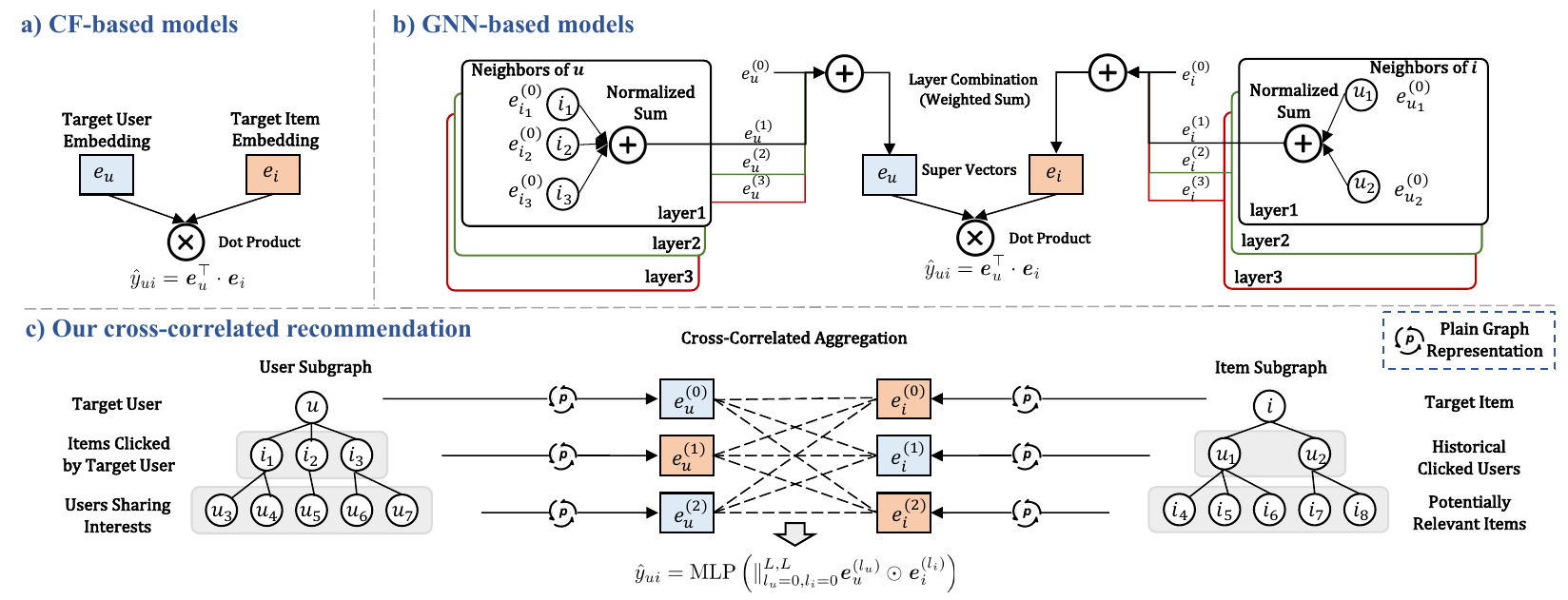}
	\caption{\textbf{(a) \& (b): Sketch of existing recommendation models; (c): The overall framework architecture of our proposed GCR.} 
    (a) CF-based recommendation models utilize the dot product of the learned user/item embeddings to infer user-item interactions. (b) GNN-based recommendation models begin by performing graph convolutions to generate embeddings for multi-hop neighbors. These embeddings are then combined to construct super vectors, which are used for making predictions. (c) GCR first extracts graph embeddings with Plain Graph Representation and then flexibly considers all potential cross-correlations between the target user and the target item with Cross-Correlated Aggregation to infer their interaction.}
	\label{fig:framework}
\end{figure*}

In order to model the intricate cross-correlations among users and items, we aim to propose a flexible framework that can effectively learn and identify dominant groups of neighbors, which should also be applicable to enhance current recommender models.
Several existing studies have explored the use of the MLP rather than the dot product~(Eq.\ref{eq:inner-product}) to implement the relevance function $r$~\cite{he2017ncf,tay2018latent}. Theoretically, an MLP is able to approximate any continuous function if the number of hidden units/layers is sufficient, according to the universal approximation theorem~\cite{csaji2001approximation}. However, formulating $r$ in this way encounters two issues. Firstly, it lacks topology awareness as it only considers the original user/item embeddings without incorporating graph embeddings like GNN-based recommendation models, which could impede the expressivity of modeling multi-hop user-item relevance. Second, although MLPs are theoretically universal, they are difficult to use to approximate the dot product in practice~\cite{rendle2020ncfmf} and thereby exhibit inferior performance than specially-designed architectures~\cite{rendle2010factorization,guo2017deepfm}.

To address the aforementioned two issues, we propose a novel simple but effective framework - Graph Cross-correlated Network for Recommendation~(GCR). The core idea of GCR is illustrated in~\autoref{fig:framework}-(c). It consists of two major components: Plain Graph Representation (PGR) and Cross-Correlated Aggregation (CCA). PGR encodes the interaction topology in an efficient way, while CCA then further flexibly aggregates the cross-correlated terms among multi-hop user/item subgraph embeddings. We summarize the idea of our proposed GCR with the following forms, which redefine Eq.~(\ref{eq:gen}) as follows:
\begin{equation}
    \begin{aligned}
        \bm{h}_x = \text{PGR}(x;\gG, \mE_{\gU},\mE_{\gI}),\quad &x\in\{u,i\}, \\
        r_{GCR}(u,i)=\text{CCA}(\bm{h}_u, &\bm{h}_i).
    \end{aligned}
\end{equation}
We now introduce the details of the PGR and CCA components in the following subsections.

\subsection{Plain Graph Representation (PGR)}
\label{sec:pgr}
As depicted in Figure 1-(c), the Plain Graph Representation (PGR) module operates in two steps. Firstly, it extracts a subgraph of a specific depth for each user/item instance. Secondly, it performs information aggregation over the retrieved subgraph in an unfolding manner.

In particular, we iteratively track the neighbors of each target node $j$ with the breadth-first searching, obtaining $\{\gN_j^{(0)},\gN_j^{(1)},\cdots,\gN_j^{(L)}\}$ where $\gN_j^{(l)}$ contains the $l$-hop neighbors of $j$, and $\gN_j^{(0)}$ contains the node $j$ itself, i.e., $\gN_j^{(0)}={j}$. We apply the simple but effective mean-pooling operator to read out the representation for each hop of neighbors, namely as follows:
\begin{equation}
\begin{aligned}
    \ve_j^{(l)} & = \text{Readout}(\{..., \ve_k^{(0)}, ...\}), k\in\gN_j^{(l)},\\ 
    &= \frac{1}{|\gN_j^{(l)}|}\sum_{k\in\gN_j^{(l)}} \ve_k^{(0)},
\end{aligned}
\label{eq:neighbor}
\end{equation}
where $\ve_j^{(l)}$ is the readout vector of node $j$'s $l$-hop neighbors.
	
{\bf Compared with GNN-based models.} The superiority of our PGR to GNN-based ways for recommendation can be illustrated in the following three main aspects.
\textbf{Firstly}, PGR is more suitable for recommendation tasks since it can extract graph information without mixing the personalized embeddings of users and items. Recursive graph convolution in Eq.(\ref{eq:GNE}) is initially designed for homogeneous graphs, where all users are of the same kind. Passing messages along the edges helps to better extract local graph information. However, for the recommendation task in the user-item bipartite graph, users and items have totally different semantic meanings and connecting characteristics. For example, users usually have some stationary interests and click a certain range of items, while popular items may be exposed and clicked by a great number and a wide range of users. Thus, mixing the embeddings of different types of nodes may bring in information noise when modeling the relations between users and items. 
\textbf{Secondly}, PGR, even though simple, is more robust than GNN for the recommendation task.
Note that the observed connections between users/items in the recommendation are usually noisy and incomplete. Performing recursive message passing like GNNs would spread the noise along with the noisy connections, hence impeding the model training. In PGR, we use mean pooling to devise the representation for each instance, which is robust to eliminate the noise.
\textbf{Lastly}, PGR is more computationally efficient, as no recursive updating is needed, ensuring the efficiency potential for online recommendations. 
Besides, PGR is parameter-free and can thus be pre-computed and stored to save time further, enhance model efficiency, and accelerate the inference process. The experimental comparisons between PGR and GNN-like models are conducted and reported in Table~\ref{tab:pgr} and Table~\ref{tab:ablation} in Section \ref{sec:exper}.

\subsection{Cross-Correlated Aggregation (CCA)}
The success of FM~\cite{rendle2010factorization,guo2017deepfm} indicates that, for recommendation tasks, employing a specially designed feature interaction model structure can greatly enhance recommendation performance compared to blindly feeding all features into an MLP structure. Earily, NIA-GCN~\cite{sun2020niagcn} first conducts interaction between a central node and its neighbors, which shows the effectiveness of building correlations. However, there is still much to be done to study how to model the complex correlations underlying user/item subgraphs for meaningful information mining.

Building upon the practical physical semantics of multi-hop embeddings, several insights can be derived.
For example, start from the view of the root target user $u$, $\ve_u^{(0)\top} \cdot\ve_i^{(0)}$ infers the explicit relation between the target user and the target item, $\ve_u^{(0)\top} \cdot\ve_i^{(1)}$ then reflects the correlation between the target user and the users who have clicked on the target item. Furthermore, $\ve_u^{(0)\top} \cdot\ve_i^{(2)}$ implies the correlation between the target user and relevant items that are clicked by the users who have clicked on the target item. 
The relations start from the view of the root target item $i$ with $\ve_i^{(0)}$ also have their physical semantic meanings.
Since different cross-correlations reflect user-item interactions from different angles, it is worth considering all possible cross-correlations between the target user and the target item in a flexible manner to make comprehensive recommendations. Thus, we provide two flexible ways for cross-correlation modeling: \textit{Hop-level Cross-Correlation} and \textit{Element-level Cross-Correlation} modeling.
	
{\bf Hop-level Cross-Correlation (HCC).} One simple but powerful way to consider cross-correlations is to model them using the dot product. Specifically, we denote the cross-correlated relevance between the $l_u$-th hop of the user graph embeddings and the $l_i$-th hop of the item embeddings as $z_{ui}^{(l_u, l_i)}$. The cross-correlated terms of HCC can be formally defined as:
\begin{equation}
\label{eq:mcc}
z_{ui}^{(l_u, l_i)} = \ve_u^{(l_u)} \cdot \ve_i^{(l_i)},
\end{equation}
where $0\leq l_u\leq L$ and $0\leq l_i\leq L$. By concatenating all possible pairs into a vector, the aggregated cross-correlation of the target user $u$ and the target item $i$ can be reflected by concatenating the cross-correlations of all possible pairs as $\parallel_{l_u=0,l_i=0}^{L,L}\vz_{ui}^{(l_u, l_i)}$. After aggregating the cross-correlations, the relevance score computation can be given as:
\begin{equation}
\begin{aligned}
    \label{eq:mcccross}
    \hat{y}_{ui} & = \text{MLP}(\parallel_{l_u=0,l_i=0}^{L,L}\vz_{ui}^{(l_u, l_i)}) \\
    & = \text{MLP}(\parallel_{l_u=0,l_i=0}^{L,L} \ve_u^{(l_u)} \cdot \ve_i^{(l_i)}).
\end{aligned}
\end{equation}
In HCC, the dimension of the cross-correlated terms $z_{ui}^{(l_u, l_i)}$ decreases to one. Note that the aggregated cross-correlation vector of HCC contains only $(L+1)^2$ scalars, meaning that the parameters of the MLP in Eq.~(\ref{eq:mcccross}) will be considerably small. This will potentially avoid the risk of overfitting and contain more information in fewer scalars.
	
{\bf Element-level Cross-Correlation (ECC).} Another way to represent the cross-correlations between multi-hop user/item graph embeddings is by computing the element-wise product instead of the dot product to preserve more latent space information. Specifically, we refer to this as the Element-level Cross-Correlation~(ECC). The computation of ECC cross-correlated terms can be expressed as follows:
\begin{equation}
\label{eq:elm}
\vz_{ui}^{(l_u, l_i)} = \ve_u^{(l_u)}\odot\ve_i^{(l_i)},
\end{equation}
where $\odot$ denotes the element-wise multiplication (Hadamard product), $0\leq l_u\leq L$ and $0\leq l_i\leq L$. After traversing all possible pairs, concatenating their element-wise multiplication into a vector, and feeding the cross-correlated terms into Eq.~(\ref{eq:mcccross}), we can arrive at:
\begin{equation}
\begin{aligned}
    \label{eq:ecccross}
    \hat{y}_{ui} & = \text{MLP}(\parallel_{l_u=0,l_i=0}^{L,L} \vz_{ui}^{(l_u, l_i)}) \\
    & = \text{MLP}(\parallel_{l_u=0,l_i=0}^{L,L}\ve_u^{(l_u)}\odot\ve_i^{(l_i)}).
\end{aligned}
\end{equation}
In ECC, the dimension of the cross-correlated terms $z_{ui}^{(l_u, l_i)}$ is the same as the dimensions of the user/item embeddings, denoted by $d$. Thus, the total dimension of the aggregated cross-correlation vector for ECC is $(L+1)^2\cdot d$.

{\bf Theorem 1} (Degree of freedom). In statistics, The concept of degrees of freedom refers to the number of values within a statistic that are allowed to vary freely. When estimating statistical parameters, the degrees of freedom indicate the amount of independent information or data used in the estimation. It is calculated by subtracting the number of parameters used as intermediate steps in the estimation from the number of independent scores contributing to the estimate.
Mathematically, degrees of freedom correspond to the number of dimensions in the domain of a random vector. It signifies the number of "free" components within the vector, or rather, the number of components that need to be known before the vector is fully determined.
	
{\bf Analysis of Flexibility.} Both HCC and ECC exhibit remarkably higher degrees of freedom than GNN-based recommendation models. 
For instance, as shown in Eq.(\ref{eq:NGCF}), the concatenating aggregation used in NGCF~\cite{wang2019ngcf} can be seen as computing the summation of the dot product for each hop of user/item graph embeddings. The concatenating aggregation cannot provide flexibility in aggregating the cross-correlation of multi-hop user-item interactions. For LightGCN~\cite{he2020lightgcn}, as illustrated in Eq.(\ref{eq:LGN}), the weighted summation aggregation of the cross-correlation is controlled by $2\cdot(L+1)$ parameters, resulting in the flexibility of $2\cdot(L+1)$. In the case of GCR, MLP structures are trained to flexibly aggregate the cross-correlations between users and items. 
The degrees of freedom for HCC and ECC can be computed as $(L+1)^2\cdot H_n^{H_l}$, where $H_l$ is the number of hidden layers in MLP, and $H_n$ is the number of hidden units of each layer. Specifically, given $L=2$, $d=64$, $H_n=256$, and $H_l=1$, the degrees of freedom for NGCF, LightGCN, HCC, and ECC are 0, 6, 2304, and 147456, respectively. Moreover, even when substituting the MLP used in GCR with linear regression, the degrees of freedom for HCC and ECC are 9 and 576, respectively, which are still higher than those of GNN-based models. In general, both theoretical and quantitative analyses demonstrate that GCR models have significantly higher degrees of freedom than GNN-based models.

Note that the cross-correlation aggregation, that is applied before MLP in Eq.~(\ref{eq:mcccross}) and Eq.~(\ref{eq:ecccross}), is crucial as it integrates the explicit user-item interactions as well as the implicit correlations between user-item pairs. We demonstrate the necessity of CCA in our ablation study and discuss the weight parameters assigned to each cross-correlated term in the case study in the following section.

\subsection{Optimization}
In our framework, PGR is parameter-free, and the parameters of MLP in Eq.(\ref{eq:mcccross}) and Eq.(\ref{eq:ecccross}) are required to be optimized. To do so, we employ the Bayesian Personalized Ranking~(BPR) loss~\cite{rendle2012bprmf,wang2019ngcf,he2020lightgcn}, which has been extensively used in recommender systems. It considers the relative order between observed and unobserved user-item interactions. Specifically, BPR assumes that the observed interactions should be assigned higher prediction values than unobserved ones. The objective function is formulated as follows:
\begin{equation}
\label{eq:bpr}
\gL_{BPR} = -\sum_{(u,i,j) \in \mathcal{O}} \ln \sigma(\hat{y}_{ui} - \hat{y}_{uj}),
\end{equation}
where $\mathcal{O} = \{(u,i,j)| (u,i)\in \mathcal{R}^{+},(u,j) \in \mathcal{R}^{-}\}$, $\mathcal{R}^+$ indicates observed user-item interactions, $\mathcal{R}^-$ denotes sampled negative interactions, and $\sigma(\cdot)$ is the Sigmoid function.

\textbf{Flexible optimization}. The components of the proposed GCR framework are task-agnostic, thus it is flexible to different recommendation tasks. For the critical CTR prediction task, we can replace the objective function with the binary cross-entropy (BCE) loss~\cite{guo2017deepfm,zhou2018din,zhou2019dien}.
Formally, for each labeled user-item pair $(u, i)$ in the training set $\gT$ of CTR prediction, the BCE objective function can be expressed as:
\begin{equation}
    \label{bce}
    \gL_{BCE} = -\sum_{(u,i)\in\gT} y_{ui}\cdot \ln(\hat{y}_{ui}) + (1-y_{ui})\cdot \ln(1-\hat{y}_{ui}),
\end{equation}
where $\hat{y}_{ui}$ is the predicted CTR and $y_{ui}$ is the ground-truth clicking label.
The overall algorithm flow of each stage of our proposed GCR for recommendation is shown in Algorithm \ref{alg:gcr-pgr} and Algorithm \ref{alg:gcr-cca}.

\begin{algorithm}[htbp] 
    \caption{Plain Graph Representation (PGR) stage of GCR} 
    \label{alg:gcr-pgr} 
    \begin{algorithmic}[1] 
        \REQUIRE The user-item interaction graph $\gG=(\gU, \gI, \mR)$; The subgraph depth $L$.
        \ENSURE The plain subgraph representations $\{\bm{E}_{\gU}^{1}, ..., \bm{E}_{\gU}^{L}\}$ and $\{\bm{E}_{\gI}^{1}, ..., \bm{E}_{\gI}^{L}\}$.
        \STATE //\textit{The process of PGR is parameter-free. Thus, we pre-compute this part in practical applications.}
        \STATE Assign users in $\gU$ with trainable embeddings $\bm{E}_{\gU}$ and items in $\gI$ with trainable embeddings $\bm{E_{\gI}}$;
        \FOR{each user/item node $j \in \gU \cup \gI$}
        \FOR{$l \in 1, ..., L$}
        \STATE Obtain $j$'s $l$-hop neighbors $\gN_{j}^{(l)}$;
        \STATE Read out the representation $\ve_{j}^{(l)}$ of $\gN_{j}^{(l)}$ via Eq.(\ref{eq:neighbor});
        \ENDFOR
        \STATE Obtain the unfolded representations $\{\ve_{j}^{(1)}, ..., \ve_{j}^{(L)}\}$ and pre-store to the database.
        \ENDFOR
    \end{algorithmic} 
\end{algorithm}

\vspace{-0.5em}
\begin{algorithm}[htbp] 
    \caption{Cross-Correlated Aggregation (CCA) stage of GCR} 
    \label{alg:gcr-cca} 
    \begin{algorithmic}[1] 
        \REQUIRE The user set $\gU$ and item set $\gI$; The embedding matrices $\bm{E}_{\gU}$ and $\bm{E}_{\gI}$; The PGR $\{\bm{E}_{\gU}^{1}, ..., \bm{E}_{\gU}^{L}\}$ and $\{\bm{E}_{\gI}^{1}, ..., \bm{E}_{\gI}^{L}\}$.
        \ENSURE The recommendation results.
        \STATE Randomly initialize the trainable parameters $\theta_{CCA}$ in the CCA module.
        \STATE //\textit{Option 1: Hop-level Cross-Correlation (HCC)}.
        \STATE //\textit{Option 2: Element-level Cross-Correlation (ECC)}.
        \IF{HCC}
        \STATE Train $\theta_{CCA}$ and the embeddings via Eq.(\ref{eq:mcc}) and Eq.(\ref{eq:mcccross}) and optimize them with recommendation task-specific objective functions (e.g., the BPR loss $\gL_{BPR}$);
        \ELSE
        \STATE Train $\theta_{CCA}$ and the embeddings via Eq.(\ref{eq:elm}) and Eq.(\ref{eq:ecccross}) and optimize them with recommendation task-specific objective functions (e.g., the BPR loss $\gL_{BPR}$);
        \ENDIF
        \STATE Conduct inference with the trained $\theta_{CCA}$ and embeddings and obtain the recommendation results.
    \end{algorithmic} 
\end{algorithm}

\section{experiments}\label{sec:exper}

We conduct extensive experiments on three datasets and an industrial dataset with the goal of answering the following five research questions:
\textbf{Q1:} Does GCR achieve the best recommendation performance among all baselines?
\textbf{Q2:} Is GCR more effective than traditional single vector dot product paradigms?
\textbf{Q3:} How efficient is GCR to be used than the traditional dot product paradigms?
\textbf{Q4:} Whether PGR extracts more information than message-passing GNN, and how do HCC and ECC affect the final recommendation performance?
\textbf{Q5:} Can GCR also be flexible and effective to equip with the CTR prediction models?

\subsection{Experimental Setup}
\subsubsection{Datasets}
We evaluate our proposed GCR on three publicly available recommendation datasets: \textbf{Gowalla}~\cite{cho2011gowalla}, \textbf{Yelp2018}~\cite{he2020lightgcn}, and \textbf{Amazon-Book}~\cite{mcauley2015image} for the commonly adopted interaction prediction recommendation task. Furthermore, an additional large-scale industrial dataset \textbf{WeiXin} is adopted for the evaluation of the CTR prediction recommendation task. 
The statistics of these datasets are shown in Table \ref{tab:stats}. 
The detailed descriptions of these datasets are shown as:
\begin{itemize}
    \item \textbf{Gowalla}\footnote{\url{https://snap.stanford.edu/data/loc-Gowalla.html}}~\cite{cho2011gowalla} is a check-in dataset provided by Gowalla with 107,092 users, 1,280,969 items and 6,442,892 check-in records, where users share their locations by checking in. In this dataset, the recorded locations are considered as items. We use this dataset to evaluate interaction prediction.
    \item \textbf{Yelp2018}\footnote{\url{https://www.kaggle.com/datasets/yelp-dataset/yelp-dataset}}~\cite{he2020lightgcn} is derived from the 2018 edition of the Yelp contest, which focuses on local businesses such as restaurants and bars. It contains 31,668 users, 38,048 items, and 1,561,406 review interactions. We use this dataset to evaluate interaction prediction.
    \item \textbf{Amazon-Book}\footnote{\url{http://jmcauley.ucsd.edu/data/amazon}}~\cite{mcauley2015image} is a subset of the Amazon dataset, which is a consumer\&book-specific recommendation dataset. It contains 52,463 users (consumers) and 91,599 items (books), with the formed 2,984,108 review interactions. We use this dataset for the evaluation of both interaction prediction and CTR prediction tasks due to it is widely adopted in both two tasks.
    \item \textbf{WeiXin} is an industrial dataset compiled from 7.3 million anonymous records of video playback logs on the Channels platform of Weixin, containing about 20,000 users and 100,000 videos. We use this dataset to evaluate CTR prediction. If a user has watched more than 90\% of a given video, we treat this record as a positive sample. Otherwise, the record will be treated as a negative sample. 
\end{itemize}

\begin{table}[tbp]
  \centering
  \small
  \caption{Statistics of the experimental datasets.}
  \vspace{-0.7em}
  \resizebox{\linewidth}{!}{
    \begin{tabular}{c|c|c|c|c|c}
    \toprule
    Dataset & \# Users & \# Items & \# Interactions &  Density & Domain \\
    \midrule
    Gowalla & 107,092 & 1,280,969 & 6,442,892 & 0.0047\% & Check-in  \\
    Yelp2018 & 31,668 & 38,048 & 1,561,406 & 0.1296\% & Business   \\
    Amazon-Book & 52,463 & 91,599 & 2,984,108 & 0.0620\% & E-commerce \\
    \midrule
    WeiXin &  $\sim$20,000 & $\sim$100,000 & $\sim$7,300,000 & 0.3650\% & Short Video \\
    \bottomrule
    \end{tabular}%
    \vspace{-0.7em}
  }
  \label{tab:stats}%
\end{table}%

\subsubsection{Comparing Methods}
We evaluate the widely adopted interaction prediction recommendation task as the main evaluation.
We compare GCR with ten representative recommendation models as follows: 
\textbf{GRMF}~\cite{rao2015grmf} is a special kind of the traditional matrix factorization method~\cite{koren2009MF}, which adds the graph Laplacian regularizer to restrict connected nodes to have similar embeddings.
\textbf{MetaPath2Vec}~\cite{dong2017metapath2vec} formalizes the metapath-based random walks on bipartite graphs as a corpus and then leverages skip-gram~\cite{mikolov2013word2vec} models to compute the node embeddings.
\textbf{NGCF}~\cite{wang2019ngcf} is a representative GNN-based model that introduces the graph collaborative filtering algorithm to model the high-order connectivity information in the embedding function.
\textbf{LightGCN}~\cite{he2020lightgcn} linearly propagates user/item information on the user-item interaction graph with the simplification of the update operation by eliminating non-linearities in NGCF and uses the weighted sum to aggregate the layer-wise embeddings as the final embeddings.
\textbf{UltraGCN}~\cite{mao2021ultragcn} is an efficient graph-based recommendation model with multi-task auxiliary losses.
\textbf{SGL}~\cite{wu2021sgl} is a self-supervised method that utilizes node dropout, edge dropout, and random walk augmentations on the user-item bipartite graph. It generates two augmented graphs with the same type of augmentation operator and employs a shared LightGCN encoder for learning user/item embeddings.
\textbf{GTN}~\cite{fan2022gtn} is a GNN-based recommendation model with the designed graph trend filtering for accuracy recommendations with robustness. 
\textbf{NCL}~\cite{lin2022ncl} is a state-of-the-art approach that employs a prototypical contrastive objective to capture user/item correlations with prototypes representing semantic neighbors. 
\textbf{SimGCL}~\cite{yu2022simgcl} is a method that introduces random uniform noise to hidden representations as an augmentation technique for graph-based recommenders. \textbf{CGCL}~\cite{he2023cgcl} is a model that enhances the graph contrastive learning with candidate views.

To further verify the flexibility of GCR for enhancing different recommendation tasks, we have also conducted experiments that equipped the CTR prediction model with GCR. The popular CTR prediction models have been introduced in the related works.
The utilized CTR prediction baselines in the experiments are listed as follows:
\textbf{DIN}~\cite{zhou2018din} is a representative model that uses a deep network based on the attention mechanism to extract user interest from historical user behaviors.
\textbf{DIEN}~\cite{zhou2019dien} is the enhanced version of DIN, which further equips the model with GRU for evolution user interest modeling.
\textbf{UBR4CTR}~\cite{qin2020ubr4ctr} and \textbf{SIM}~\cite{pi2020search} are representative CTR prediction method that aims to model the user's life-long behavior sequence with the search-based paradigm. For SIM, we use its hard search paradigm in our experiments. Further, \textbf{GMT}~\cite{min2022gmt} is a state-of-the-art graph-based CTR prediction model with neighborhood sampling and graph transformer architecture. \textbf{DCIN}~\cite{li2023dcin} is a novel interest modeling model for CTR prediction with the inclusion of decision-making contexts.

\begin{table*}[t]
\small
\centering
\caption{Overall recommendation performance of GCR compared with state-of-the-art baseline models on three experimental datasets. The best and second-best results are highlighted in \textbf{bold} font and \underline{underlined}. * indicates the statistical significance of improvement over the best-performed baseline with $p < 0.05$.}
\resizebox{0.915\linewidth}{!}{
\begin{tabular}{cccccccccc}
\toprule
    Datasets  & \multicolumn{3}{c}{Gowalla} & \multicolumn{3}{c}{Yelp2018} & \multicolumn{3}{c}{Amazon-Book} \\
\midrule
Methods & Precision & Recall & NDCG  & Precision & Recall & NDCG  & Precision & Recall & NDCG \\
\cmidrule[0.5pt](r){1-1} \cmidrule[0.5pt](r){2-4} \cmidrule[0.5pt](r){ 5-7} \cmidrule[0.5pt](r){8-10}
GRMF~\cite{rao2015grmf}  & 0.0329 & 0.1158 & 0.0926 & 0.0236 & 0.0502 & 0.0409 & 0.0173 & 0.0357 & 0.0307 \\
MetaPath2Vec~\cite{dong2017metapath2vec} & 0.0402 & 0.1415 & 0.1062 & 0.0227 & 0.0458 & 0.0411 & 0.0358 & 0.0740 & 0.0683 \\
\cmidrule[0.5pt](r){1-1} \cmidrule[0.5pt](r){2-4} \cmidrule[0.5pt](r){ 5-7} \cmidrule[0.5pt](r){8-10}
NGCF~\cite{wang2019ngcf}  & 0.0497 & 0.1570 & 0.1365 & 0.0322 & 0.0673 & 0.0569 & 0.0411 & 0.0851 & 0.0705 \\
LightGCN~\cite{he2020lightgcn} & 0.0537 & 0.1834 & 0.1422 & 0.0421 & 0.0902 & 0.0786 & 0.0459 & 0.0941 & 0.0845 \\
\textcolor{black}{UltraGCN~\cite{mao2021ultragcn}}  & \textcolor{black}{0.0543} & \textcolor{black}{0.1842} & \textcolor{black}{0.1431} & \textcolor{black}{0.0436} & \textcolor{black}{0.0923} & \textcolor{black}{0.0823} & \textcolor{black}{0.0458} & \textcolor{black}{0.0934} & \textcolor{black}{0.0839}  \\
SGL~\cite{wu2021sgl}  & 0.0554 & 0.1848 & 0.1467 & 0.0460 & 0.0951 & 0.0846 & 0.0468 & 0.0956 & 0.0875 \\
GTN~\cite{fan2022gtn}  & 0.0564 & 0.1904 & 0.1515 & 0.0429 & 0.0908 & 0.0785 & 0.0461 & \underline{0.0975} & 0.0857  \\
\textcolor{black}{NCL~\cite{lin2022ncl}}  & \textcolor{black}{0.0568} & \textcolor{black}{0.1920} & \textcolor{black}{0.1523} & \textcolor{black}{0.0457} & \textcolor{black}{0.0944} & \textcolor{black}{0.0825} & \textcolor{black}{0.0471} & \textcolor{black}{0.0970} & \textcolor{black}{0.0871}  \\
\textcolor{black}{SimGCL~\cite{yu2022simgcl}}  & \textcolor{black}{\underline{0.0579}} & \textcolor{black}{\underline{0.1932}} & \textcolor{black}{\underline{0.1543}} & \textcolor{black}{\underline{0.0463}} & \textcolor{black}{\underline{0.0968}} & \textcolor{black}{\underline{0.0858}} & \textcolor{black}{\underline{0.0474}} & \textcolor{black}{0.0973} & \textcolor{black}{\underline{0.0878}}  \\
\textcolor{black}{CGCL~\cite{he2023cgcl}}  & \textcolor{black}{0.0563} & \textcolor{black}{0.1893} & \textcolor{black}{0.1504} & \textcolor{black}{0.0460} & \textcolor{black}{0.0961} & \textcolor{black}{0.0847} & \textcolor{black}{0.0463} & \textcolor{black}{0.0962} & \textcolor{black}{0.0861}  \\
\cmidrule[0.5pt](r){1-1} \cmidrule[0.5pt](r){2-4} \cmidrule[0.5pt](r){5-7} \cmidrule[0.5pt](r){8-10}
\textbf{GCR} (ours)   & \textbf{0.0604} & \textbf{0.1982} & \textbf{0.1588} & \textbf{0.0476} & \textbf{0.0991} & \textbf{0.0867} & \textbf{0.0485} & \textbf{0.0992} & \textbf{0.0891} \\
\textcolor{black}{Improvement (\%)} & \textcolor{black}{4.32\%*} & \textcolor{black}{2.59\%*} & \textcolor{black}{2.92\%*} & \textcolor{black}{2.81\%*} & \textcolor{black}{2.38\%*} & \textcolor{black}{1.05\%*} & \textcolor{black}{2.32\%*} & \textcolor{black}{1.74\%*} & \textcolor{black}{1.48\%*} \\
\bottomrule
\end{tabular}%
}
\label{tab:result}
\end{table*}

\subsubsection{Evaluation Details}
For the main results, we randomly split the user-item interaction records of each dataset into three distinct subsets: an embedding pre-training set, a model training (validation) set, and a test set, to evaluate the performance. These subsets respectively account for 65\%, 15\%, and 20\% of the entire dataset. We use LightGCN~\cite{he2020lightgcn} to produce the pre-trained embeddings of users and items in datasets and then feed them into the models as the raw features for model training. For the CTR prediction results, assuming each user has $T$ records, we use its $[1, T-2]$ records for model training, its $(T-1)$-th record for validation, and its $T$-th record for CTR prediction testing~\cite{qin2020ubr4ctr,bei2023non}.

For evaluation metrics, we adopt the popular all-ranking evaluation protocol, which has been widely used in recent studies~\cite{wang2019ngcf,he2020lightgcn}.
For each user in the testing set, all non-interacted items are treated as negative items. Specifically, we rank all items in the dataset except for the interacted items used in the training process and then truncate the ranked list at 20 to calculate the Precision@20, Recall@20, and NDCG@20 metrics followed the previous works~\cite{wang2019ngcf,he2020lightgcn}.
Then, we have also conducted the analysis of the CTR prediction performance of GCR in subsection~\ref{sec:ctr}. The evaluation metrics utilized for CTR prediction are the widely used AUC and RelaImpr~\cite{zhou2018din}.

\subsubsection{Implementation Details}
The embedding size is fixed to 64 for all models, and all the embedding methods are implemented with their official codes. 
For our GCR, regarding the PGR module, we restrict the number of subgraph layers to 3. 
For the training of GCR, we use the Adam optimizer~\cite{kingma2015adam} with the learning rate searched from \{0.01, 0.005, 0.001, 0.0005, 0.0001\} and a batch size of 512. The coefficient of $\ell_2$ normalization is set as $1e^{-5}$, and the dropout ratio is set as 0.7. For non-linear layers, we use the Xavier distribution~\cite{jakovetic2014fast} to initialize the model parameters and utilize batch normalization~\cite{ioffe2015batch} with a momentum of 0.1 to normalize the inputs before each non-linear layer. The activation function in MLP is set as Leaky-ReLU whose negative slope is set as 0.02. The size of the hidden layers is set as 256, while the number of hidden layers of MLP is set as 1 by default. Moreover, we perform an early stopping strategy by observing the AUC scores on the evaluation set. By default, GCR represents using PGR-ECC to flexibly consider all cross-correlation terms.
Note that we ran all the experiments ten times with different random seeds and reported the average results.

\subsection{Overall Recommendation Results (Q1)}
\label{sec:results}
In this subsection, we compare our proposed GCR with the other six representative recommendation baseline models. The overall recommendation comparison results are illustrated in Table~\ref{tab:result}, of which the improvement is calculated by comparing the performance of GCR with the best-performed baseline (marked in underlined).
From these results, we have the following observations:
\begin{itemize}
    \item Firstly, it can be seen from the table that our GCR statistically outperforms the best baseline models on all datasets with respect to all metrics. For instance, GCR demonstrates a significant improvement in Precision of 4.32\%, 2.81\%, and 2.32\% across the three experimental datasets, respectively. Specifically, GCR brings the performance average gains of 3.15\%, 2.24\%, and 1.82\% for Precision, Recall, and NDCG metrics respectively on the three experimental datasets.
    \item Furthermore, SimGCL mainly obtains the best performance over all the baselines. NGCF, LightGCN, UltraGCN, SGL, GTN, NCL, and SimGCL consistently outperform GRMF and MetaPath2Vec, suggesting that considering and aggregating the graph information helps to provide better recommendation results. GCR provides a more flexible way to aggregate the graph information and thus achieves better performance.
\end{itemize}

\begin{table*}[thbp]
\small
\centering
\caption{Comparison of different relevance functions over three embeddings and three experimental datasets. * indicates the statistical significance of improvement over the best-performed base relevance function with $p < 0.05$.}
\resizebox{\linewidth}{!}{
\begin{tabular}{ccccccccccc}
\toprule
      &       & \multicolumn{3}{c}{Gowalla} & \multicolumn{3}{c}{Yelp2018} & \multicolumn{3}{c}{Amazon-Book} \\
\cmidrule[0.5pt](r){1-1} \cmidrule[0.5pt](r){2-2}\cmidrule[0.5pt](r){3-5} \cmidrule[0.5pt](r){6-8} \cmidrule[0.5pt](r){9-11}
Embeddings & Relevance Function & Precision & Recall & NDCG  & Precision & Recall & NDCG  & Precision & Recall & NDCG \\
\cmidrule[0.5pt](r){1-1} \cmidrule[0.5pt](r){2-2}\cmidrule[0.5pt](r){3-5} \cmidrule[0.5pt](r){6-8} \cmidrule[0.5pt](r){9-11}
\multirow{5}[0]{*}{GRMF} & Dot Product & 0.0329 & 0.1158 & 0.0926 & 0.0236 & 0.0502 & 0.0409 & 0.0173 & 0.0357 & 0.0307 \\
      & MLP   & 0.0281 & 0.0970 & 0.0756 & 0.0236 & 0.0512 & 0.0415 & 0.0158 & 0.0334 & 0.0277 \\
      & Super-Vector   & \underline{0.0331} & \underline{0.1165} & \underline{0.0935} & \underline{0.0255} & \underline{0.0537} & \underline{0.0437} & \underline{0.0190} & \underline{0.0388} & \underline{0.0335} \\
      \cmidrule[0.5pt](r){2-2}\cmidrule[0.5pt](r){3-5} \cmidrule[0.5pt](r){6-8} \cmidrule[0.5pt](r){9-11}
      & \textbf{GCR}   & \textbf{0.0354} & \textbf{0.1227} & \textbf{0.0974} & \textbf{0.0264} & \textbf{0.0564} & \textbf{0.0453} & \textbf{0.0215} & \textbf{0.0436} & \textbf{0.0366} \\
 \cmidrule[0.5pt](r){2-2}\cmidrule[0.5pt](r){3-5} \cmidrule[0.5pt](r){6-8} \cmidrule[0.5pt](r){9-11}
      & Improvement (\%) & 6.95\%* & 5.32\%* & 4.17\%* & 3.53\%* & 5.03\%* & 3.66\%* & 13.16\%* & 12.37\%* & 9.25\%* \\
      \cmidrule[0.5pt](r){1-1} \cmidrule[0.5pt](r){2-2}\cmidrule[0.5pt](r){3-5} \cmidrule[0.5pt](r){6-8} \cmidrule[0.5pt](r){9-11}
\multirow{5}[0]{*}{MetaPath2Vec} & Dot Product & 0.0402 & 0.1415 & 0.1062 & 0.0286 & 0.0458 & 0.0411 & 0.0358 & 0.0740 & 0.0683 \\
      & MLP   & 0.0331 & 0.1065 & 0.0841 & 0.0255 & 0.0610 & 0.0494 & 0.0250 & 0.0503 & 0.0436 \\
      & Super-Vector   & \underline{0.0473} & \underline{0.1624} & \underline{0.1240} & \underline{0.0391} & \underline{0.0845} & \underline{0.0712} & \underline{0.0448} & \underline{0.0926} & \underline{0.0838} \\
      \cmidrule[0.5pt](r){2-2}\cmidrule[0.5pt](r){3-5} \cmidrule[0.5pt](r){6-8} \cmidrule[0.5pt](r){9-11}
      & \textbf{GCR}   & \textbf{0.0534} & \textbf{0.1802} & \textbf{0.1445} & \textbf{0.0433} & \textbf{0.0942} & \textbf{0.0811} & \textbf{0.0505} & \textbf{0.1012} & \textbf{0.0928} \\
       \cmidrule[0.5pt](r){2-2}\cmidrule[0.5pt](r){3-5} \cmidrule[0.5pt](r){6-8} \cmidrule[0.5pt](r){9-11}
      & Improvement (\%) & 12.90\%* & 10.96\%* & 16.53\%* & 10.74\%* & 11.48\%* & 13.90\%* & 12.72\%* & 9.29\%* & 10.74\%* \\
      \cmidrule[0.5pt](r){1-1} \cmidrule[0.5pt](r){2-2}\cmidrule[0.5pt](r){3-5} \cmidrule[0.5pt](r){6-8} \cmidrule[0.5pt](r){9-11}
\multirow{5}[0]{*}{LightGCN} & Dot Product & \underline{0.0564} & \underline{0.1904} & \underline{0.1515} & \underline{0.0421} & \underline{0.0902} & \underline{0.0786} & \underline{0.0459} & \underline{0.0941} & \underline{0.0845} \\
      & MLP   & 0.0322 & 0.1065 & 0.0827 & 0.0297 & 0.0614 & 0.0528 & 0.0226 & 0.0452 & 0.0391 \\
      & Super-Vector   & 0.0539 & 0.1825 & 0.1437 & 0.0411 & 0.0877 & 0.0757 & 0.0452 & 0.0921 & 0.0827 \\
      \cmidrule[0.5pt](r){2-2}\cmidrule[0.5pt](r){3-5} \cmidrule[0.5pt](r){6-8} \cmidrule[0.5pt](r){9-11}
      & \textbf{GCR}   & \textbf{0.0604} & \textbf{0.1982} & \textbf{0.1588} & \textbf{0.0476} & \textbf{0.0991} & \textbf{0.0867} & \textbf{0.0485} & \textbf{0.0992} & \textbf{0.0891}     \\
       \cmidrule[0.5pt](r){2-2}\cmidrule[0.5pt](r){3-5} \cmidrule[0.5pt](r){6-8} \cmidrule[0.5pt](r){9-11}
      & Improvement (\%) & 7.09\%* & 4.10\%* & 4.82\%* & 13.06\%* & 9.87\%* & 10.31\%* & 5.66\%* & 5.42\%* & 5.44\%* \\
      \bottomrule
\end{tabular}%
}
\label{tab:pgr}
\end{table*}

\subsection{Effectiveness Study of GCR (Q2)}
\label{sec:efficiency}
In this subsection, we compare the recommendation performance between our GCR and the \textit{dot product, MLP, and the super-vector-based dot product} when fed with the same embeddings. The compared relevance functions are specialized as follows:
\begin{itemize}
\item \textbf{Dot product}~\cite{koren2009MF,rendle2012bprmf}: This is the most common relevance metric given by Eq.(\ref{eq:inner-product}).
\item \textbf{MLP} (Multi layer perception)~\cite{he2017ncf,fan2019meirec}: MLP is used in NCF~\cite{he2017ncf} and common CTR recommendation models~\cite{guo2017deepfm}, which also become a widely used relevance function.
\item \textbf{Super-Vector} (Super-vector-based dot product)~\cite{wang2019ngcf,he2020lightgcn}: We use LightGCN to calculate the super vectors for each user and item.
\end{itemize}

The Effectiveness analysis is shown in Table~\ref{tab:pgr}, which provides the comparisons of the relevance functions over three different embeddings on three experimental datasets. The improvement is calculated by comparing GCR with the other best-performed relevance functions. From this result table, we can derive observations as follows.
\begin{itemize}
    \item Simply applying MLP as the relevance function cannot amplify the expressive power of the learned embedding. As stated in~\cite{rendle2020ncfmf}, although MLP is theoretically universal to approximate the dot product, it is difficult to approximate in practice, especially when the embeddings are optimized for the dot product. 
    \item Super-vector-based dot products can enhance the expressive power of GRMF and MetaPath2Vec embeddings. But for LightGCN, super-vector-based dot products lead to relatively worse recommendation performance. One possible reason for the degradations is that LightGCN has already utilized graph convolution layers to encode the graph information explicitly. Further applying graph convolution to graph-based embeddings will encounter the problem of high-order GNNs, e.g., the over-smoothing problem~\cite{chen2020measuring}. The over-smoothed embeddings are indistinguishable and negatively affect user modeling and the recommendation performance~\cite{li2018laplacian}. 
    \item \textcolor{black}{GCR statistically outperforms the other three counterparts on all three datasets with all three embeddings. In particular, with MetaPath2Vec embeddings, GCR improves over the strongest baselines \textit{w.r.t.} NDCG@20 metric by 16.53\%, 13.90\%, and 10.74\% on Gowalla, Yelp2018, and Amazon-Book, respectively.
Further, GCR outperforms the other relevant functions with average gains of 9.53\%, 8.20\%, and 8.76\% for Precision, Recall, and NDCG over the three types of embeddings on the three recommendation datasets.}
This phenomenon illustrates that GCR is able to alleviate the over-smoothing problem and further improve the recommendation performance of GRMF, MetaPath2Vec, and LightGCN.
 \end{itemize}

\begin{figure}[tbp]
\centering	
\includegraphics[width=0.8\linewidth, trim=0cm 0cm 0cm 0cm,clip]{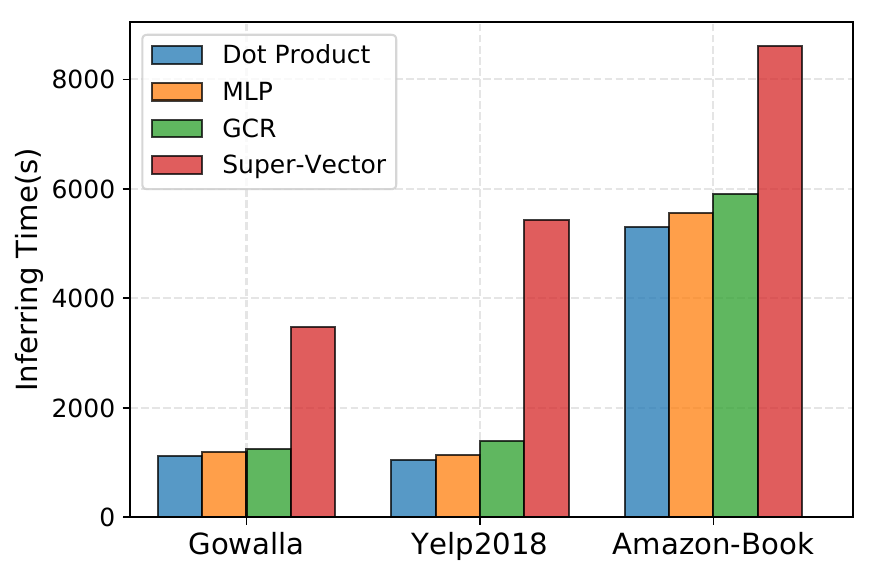}
\caption{Comparison of inferring times on three experimental datasets.}
\label{fig:infer}
\end{figure}

\subsection{Efficiency Study of GCR (Q3)}
Due to the necessity of real-time recommendation, besides effectiveness, model efficiency is also an important factor that needs to be considered~\cite{bei2023non}. As for the efficiency analysis, we report the average inference time of each compared method in \autoref{fig:infer}. 

From these results, we can see the following findings: 
First, GCR has a negligible time cost compared to the dot product and MLP, which is acceptable given the dramatic performance improvement we achieved. 
Then, compared to the super-vector-based dot product, GCR is much more efficient because PGR is a parameter-free and parallelizable process. This experiment demonstrates that although GCR extracts and exploits the cross-correlations between multi-hop user/item subgraph embeddings, it achieves a good balance between time consumption and performance.

\subsection{Ablation Study~(Q4)}
In the ablation study section, we discuss the influence of our two core components: PGR and CCA. Specifically, we compare the recommendation performance of the following six combinations:
\begin{itemize}
    \item GNN (Super-Vector): This variant uses GNN to compute the graph representation super vector and uses the dot product for interaction prediction. 
    \item GNN-HCC: This variant uses GNN for graph representation and HCC for interaction prediction. 
    \item GNN-ECC: This variant uses GNN for graph representation and ECC for interaction prediction.
     \item PGR-MLP: This variant uses PGR for graph representation and then feeds the concatenated graph representation into a three-layer MLP for interaction prediction. 
    \item PGR-HCC: This variant uses PGR for graph representation and HCC for interaction prediction. 
    \item GCR (PGR-ECC): This variant uses PGR for graph representation and ECC for interaction prediction.
\end{itemize}

\begin{table}[t]
\centering
\caption{Comparison of different GCR variations on MetaPath2Vec.}
\vspace{-0.5em}
\resizebox{\linewidth}{!}{
 \begin{tabular}{cccc}
 \toprule
  & \multicolumn{1}{c}{Gowalla} & \multicolumn{1}{c}{Yelp2018} & \multicolumn{1}{c}{Amazon-Book} \\
\midrule
   Relevance Functions  & NDCG  & NDCG  & NDCG \\
   \midrule
 GNN~(Super-Vector)   & 0.1240 & 0.0712  & 0.0838 \\
 GNN-HCC & 0.1361   & 0.0738   & 0.0853  \\
 GNN-ECC  & 0.1371    & 0.0742   & 0.0860  \\
 \midrule
 PGR-MLP & 0.0971  & 0.0541  & 0.0479 \\
 PGR-HCC & \underline{0.1438}  & \underline{0.0791} & \underline{0.0926} \\
 GCR~(PGR-ECC)  & \textbf{0.1445}  & \textbf{0.0811}  & \textbf{0.0928} \\
 \bottomrule
 \end{tabular}%
\label{tab:ablation}
}
\end{table}

\begin{figure}[!t]
\centering	
\includegraphics[width=\linewidth]{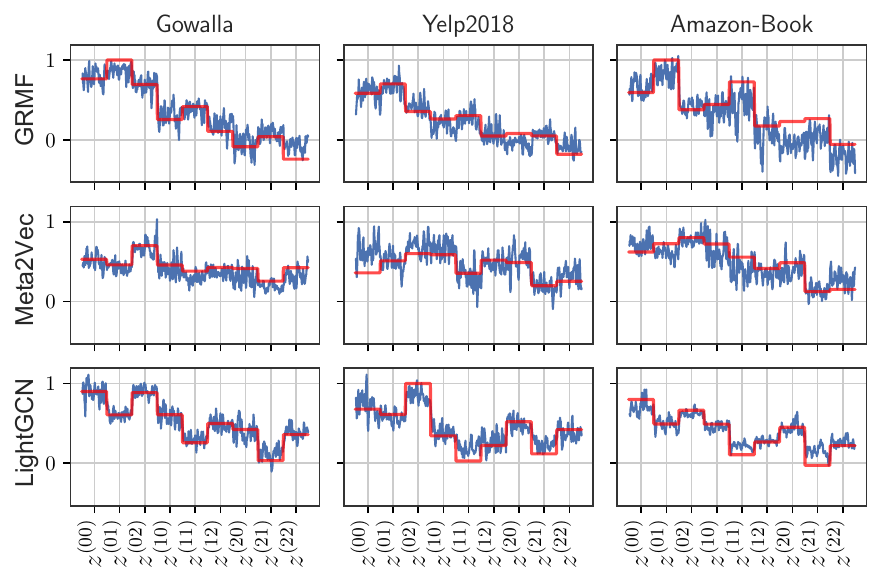}
\vspace{-0.5em}
\caption{Visualization of the weights of the cross-correlation terms. For HCC, we have single-weight parameters for these terms~(red lines), while we have 64 weight parameters for these terms of ECC~(blue lines).}
\vspace{-0.5em}
\label{fig:weight}
\end{figure}

We describe the comparison results in~\autoref{tab:ablation}, from which we present the NDCG@20 of the above six variants on Gowalla, Yelp2018, and Amazon-Book. Without loss of generality, we conduct this experiment with MetaPath2Vec embeddings. From this result table, we can summarize the following conclusions:
\begin{itemize}
    \item PGR-based variations consistently perform better than GNN-based variations. This demonstrates that PGR is more informative than GNN for recommendation tasks. The plain representation style better extracts the information of each hop of neighbors than recursively message-passing GNNs. Therefore works better with CCA to extract the implicit relation behind user and item subgraphs. 
    \item Compared to GNN~(Super-Vector), GNN-HCC/ECC consistently outperforms GNN in all three datasets. This proves that GNN~(Super-Vector), which represents users/items as super-vectors, cannot fully exploit the expressive power of embeddings. Thus, by flexibly considering all possible cross-correlations with CCA, GNN-HCC/ECC outperforms GNN~(Super-Vector) by significant margins. 
    \item PGR-MLP performs the worst on all three datasets. Without explicitly extracting the cross-correlations with CCA, MLP cannot model the complex interaction between user and item subgraphs. This observation verifies and substantiates the importance of cross-correlated aggregation.
\end{itemize}

\textbf{Case Study}. We further explore the mechanism of HCC/ECC of GCR through visualization.
Specifically, in~\autoref{fig:weight}, we visualize the weights of linear HCC/ECC by setting the number of hidden layers for MLP to zero. By visualizing the weights associated with each cross-correlated term, this figure demonstrates the following three mai observations: 1) the necessity of the cross-correlation aggregation; 2) the consistency between HCC and ECC; 3) the characteristics of the weights associated with each cross-correlated term. 

\begin{itemize}
    \item {\bf The necessity of the cross-correlated aggregation.} From the individual subplots in~\autoref{fig:weight}, it appears that CCA models, including HCC and ECC, learn different weights for each cross-correlated term. This verifies that assigning different weights for each cross-correlated term helps to reduce the personalized ranking loss as well as make more accurate recommendations. 
    \item {\bf The consistency between HCC and ECC.} One concern about the cross-correlated aggregation is whether the element-wise ECC tends to assign similar weight parameters as the layer-wise HCC model. As shown in the figure, after normalization, the weight parameters of HCC and ECC show similar trends on all three datasets with all three embeddings, confirming the consistency between HCC and ECC.
    \item {\bf The characteristic of the embeddings} In~\autoref{fig:weight}, each row of subplots represents the visualization of the weights of a particular embedding, from GRMF to MetaPath2Vec to LightGCN. Comparing the tendency of the weight parameters of each embedding, the tendency of a particular embedding seems to be similar across different datasets. For GRMF, the weight parameters of $z^{(01)}$ are significantly larger than other cross-correlation terms. For MetaPath2Vec, the weight parameters of each cross-correlation term happen to be much more uniform than other embeddings, while the weight parameters of $z^{(02)}$ are slightly higher than the others. For LightGCN, the SOTA recommendation method, $z^{(00)}$ and $z^{(02)}$ have higher importance than other cross-correlation terms.
\end{itemize}
    
\subsection{CTR Prediction Performance (Q5)}\label{sec:ctr}
Click-Through Rate~(CTR) prediction task is to predict the probability that a target user will click on a target item. 
It performs as an important stage in the recommendation system.
Therefore, in this subsection, we have also verified the flexibility and effectiveness of GCR in equipping the CTR models for performing the CTR prediction task. 

\textcolor{black}{We test whether GCR can improve the CTR prediction task by running experiments on user-item clicking data. Specifically, we conduct CTR prediction experiments on the widely used Amazon-Book recommendation dataset and the WeiXin industrial dataset, and the details of these datasets are illustrated in Table~\ref{tab:stats}.
We include six representative CTR prediction models: DIN ~\cite{zhou2018din}, DIEN~\cite{zhou2019dien}, UBR4CTR~\cite{qin2020ubr4ctr}, SIM~\cite{pi2020search}, GMT~\cite{min2022gmt}, and DCIN~\cite{li2023dcin} as our baselines. Further, we equip DIN with our GCR as the model ''DIN+GCR'', which refers to enhancing DIN with the modeling of high-order information (1-hop and 2-hop neighbors with DIN-based neighborhood sampling) and graph cross-correlation terms with our GCR model.}

\begin{table}[tbp]
  \centering
  \caption{CTR prediction results on Amazon-Book and WeiXin datasets. * indicates the statistical significance of improvement over the best-performed baseline with $p < 0.05$.}
  \resizebox{\columnwidth}{!}{
    \begin{tabular}{c|cc|cc}
    \toprule
    \multirow{2}[2]{*}{Model} & \multicolumn{2}{c|}{Amazon-Book} & \multicolumn{2}{c}{WeiXin} \\
          & AUC   & RelaImpr & AUC   & RelaImpr \\
    \midrule
    DIN~\cite{zhou2018din}  & 0.8106 & 0.00\% & 0.8071 & 0.00\% \\
    DIEN~\cite{zhou2019dien}  & 0.8218 & 3.61\% & 0.8104 & 1.07\% \\
    \textcolor{black}{UBR4CTR~\cite{qin2020ubr4ctr}} & \textcolor{black}{0.8002} & \textcolor{black}{-3.35\%} & \textcolor{black}{0.7994} & \textcolor{black}{-2.51\%} \\
    SIM~\cite{pi2020search} & 0.8304 & 6.37\% & 0.8151 & 2.61\% \\
    \textcolor{black}{GMT~\cite{min2022gmt}}   & \textcolor{black}{\underline{0.8445}} & \textcolor{black}{\underline{10.91\%}} & \textcolor{black}{\underline{0.8175}} & \textcolor{black}{\underline{3.39\%}} \\
    \textcolor{black}{DCIN~\cite{li2023dcin}}   & \textcolor{black}{0.8372} & \textcolor{black}{8.56\%} & \textcolor{black}{0.8162} & \textcolor{black}{2.96\%} \\
    \midrule
    DIN+GCR & \textbf{0.8681} & \textbf{18.51\%} & \textbf{0.8277} & \textbf{6.71\%} \\
    \textcolor{black}{Improvement (\%)} & \textcolor{black}{2.79\%*} & \textcolor{black}{--} & \textcolor{black}{1.25\%*} & \textcolor{black}{--} \\
    \bottomrule
    \end{tabular}%
    }
  \label{tab:ctr}%
\end{table}%

\textcolor{black}{In~\autoref{tab:ctr}, we present the AUC metric and corresponding RelaImpr value to evaluate the performance of these CTR prediction models. From this table, ``DIN+GCR'' outperforms the other baselines, confirming the effectiveness and flexibility of GCR on the CTR prediction task. Compared to the best baseline GMT, ``DIN+GCR'' shows a relative improvement of 2.79\% on Amazon-Book and 1.25\% on WeiXin. More importantly, ``DIN+GCR'' improves DIN by extracting the graph information with PGR and constructing the cross-correlation interactions with ECC, bringing a RelaImpr gain of 18.51\% on Amazon-Book and 6.71\% on WeiXin. 
These results also show the effectiveness of GCR for real-world use in industrial scenarios.
Thus, it can be utilized flexibly to enhance current CTR prediction models.}

\section{Conclusion}

In this paper, we study the problem of recommendation with the modeling of graph neural networks (GNNs). Existing GNNs for recommendation encode the user-item subgraph into a single representation vector and conduct inference with the dot product-based operator. However, in this manner, the physical semantic meaning of each hop of the neighborhood is severely weakened.
To better exploit the semantic information and make full use of the correlations between the information of central nodes and their neighborhoods, in this paper, we propose the graph cross-correlated network for recommendation (GCR), which is an effective framework that can flexibly model the cross-interactions among users and different-hop neighbors of items. We introduce the architecture of GCR with plain graph representation (PGR) and Cross-correlated Aggregation (CCA) components in detail, and then we theoretically analyze GCR and validate that it contains more flexibility than existing recommender models. 
Extensive experiments over three benchmark recommendation datasets and an industrial dataset demonstrate the effectiveness of GCR. Further in-depth analysis of GCR more specifically verified the effectiveness, efficiency, and flexibility of GCR.
Future works can be conducted to study more powerful graph representation methods in an unfolding plain manner and explore the potentiality of GCR for explainable recommendations.

%

\section*{Acknowledgments}
This work was supported in part by the National Natural Science Foundation of China (No. U22A2095, 62032020, 62272200) and the Innovation and Technology Commission of Hong Kong under Innovation and Technology Fund - Mainland-Hong Kong Joint Funding Scheme (MHP/012/21).


\ifCLASSOPTIONcaptionsoff
  \newpage
\fi

\bibliographystyle{IEEEtran}
\bibliography{IEEEabrv,reference_new}


\end{document}